\documentclass[12pt]{article}
\usepackage{amssymb}
\usepackage{amsmath}

\begin{document}

\begin{titlepage}
\title{\bf Lagrangian Mechanics on the standard Cliffordian K\"{a}hler Manifolds}
\author{ Mehmet Tekkoyun \footnote{Corresponding author. E-mail address: tekkoyun@pau.edu.tr; Tel: +902582953616; Fax: +902582953593}\\
{\small Department of Mathematics, Pamukkale University,}\\
{\small 20070 Denizli, Turkey}}
\date{\today}
\maketitle

\begin{abstract}

This study presents standard Cliffordian K\"{a}hler analogue of
Lagrangian mechanics. Also, the some geometric and physical
results related to the standard Cliffordian K\"{a}hler dynamical
systems are given.

{\bf Keywords:} Cliffordian K\"{a}hler geometry, Lagrangian
Mechanics.

{\bf PACS:} 02.40

\end{abstract}
\end{titlepage}

\section{Introduction}

Modern differential geometry explains explicitly the dynamics of
Lagrangians. So, we say that if $M$ is an $m$-dimensional configuration
manifold and $L:TM\rightarrow \mathbf{R}$\textbf{\ }is a regular Lagrangian
function, then there is a unique vector field $\xi $ on $TM$ such that
dynamics equations is given by
\begin{equation}
i_{\xi }\Phi _{L}=dE_{L}  \label{1.1}
\end{equation}%
where $\Phi _{L}$ indicates the symplectic form. The triple $(TM,\Phi
_{L},\xi )$ is called \textit{Lagrangian system} on the tangent bundle $TM$ $%
.$

In literature, there are a lot of studies about Lagrangian mechanics,
formalisms, systems and equations \cite{deleon, tekkoyun} and there in.
There are real, complex, paracomplex and other analogues. It is possible to
produce different analogous in different spaces. Finding new dynamics
equations is both a new expansion and contribution to science to explain
physical events.

Quaternions were invented by Sir William Rowan Hamilton as an extension to
the complex numbers. Hamilton's defining relation is most succinctly written
as:

\begin{equation}
i^{2}=j^{2}=k^{2}=ijk=-1  \label{1.2}
\end{equation}%
If it is compared to the calculus of vectors, quaternions have slipped into
the realm of obscurity. They do however still find use in the computation of
rotations. A lot of physical laws in classical, relativistic, and quantum
mechanics can be written pleasantly by means of quaternions. Some physicists
hope they will find deeper understanding of the universe by restating basic
principles in terms of quaternion algebra. It is well-known that quaternions
are useful for representing rotations in both quantum and classical
mechanics \cite{dan} . Cliffordian manifold is a quaternion manifold. The
above properties yield also for Cliffordian manifold.

In this paper, Euler-Lagrange equations related to Lagrangian systems on
Cliffordian K\"{a}hler manifold have been obtained.

\section{Preliminaries}

Throughout this paper, all mathematical objects and mappings are assumed to
be smooth, i.e. infinitely differentiable and Einstein convention of
summarizing is adopted. $\mathcal{F}(M)$, $\chi (M)$ and $\Lambda ^{1}(M)$
denote the set of functions on $M$, the set of vector fields on $M$ and the
set of 1-forms on $M$, respectively.

\subsection{Cliffordian K\"{a}hler Manifolds}

Here, we recalled the main concepts and structures given in \cite{yano,
burdujan} . Let $M$ be a real smooth manifold of dimension $m.$ Suppose that
there is a 6-dimensional vector bundle $V$ consisting of $F_{i}(i=1,2,...,6)$
tensors of type (1,1) over $M.$ Such a local basis $\{F_{1},F_{2},...,F_{6}\}
$ is called a canonical local basis of the bundle $V$ in a neighborhood $U$
of $M$. Then $V$ is called an almost Cliffordian structure in $M$. The pair $%
(M,V)$ is named an almost Cliffordian manifold with $V$. Hence, an almost
Cliffordian manifold $M$ is of dimension $m=8n.$ If there exists on $(M,V)$
a global basis $\{F_{1},F_{2},...,F_{6}\},$ then $(M,V)$ is said to be an
almost Cliffordian manifold; the basis $\{F_{1},F_{2},...,F_{6}\}$ is called
a global basis for $V$.

An almost Cliffordian connection on the almost Cliffordian manifold $(M,V)$
is a linear connection $\nabla $ on $M$ which preserves by parallel
transport the vector bundle $V$. This means that if $\Phi $ is a
cross-section (local-global) of the bundle $V$, then $\nabla _{X}\Phi $ is
also a cross-section (local-global, respectively) of $V$, $X$ being an
arbitrary vector field of $M$.

If for any canonical basis $\{J_{1},J_{2},...,J_{6}\}$ of $V$ in a
coordinate neighborhood $U$, the identities
\begin{equation}
g(J_{i}X,J_{i}Y)=g(X,Y),\text{ }\forall X,Y\in \chi (M),\text{ }\
i=1,2,...,6,  \label{2.2}
\end{equation}%
hold, the triple $(M,g,V)$ is named an almost Cliffordian Hermitian manifold
or metric Cliffordian manifold denoting by $V$ an almost Cliffordian
structure $V$ and by $g$ a Riemannian metric and by $(g,V)$ an almost
Cliffordian metric structure$.$

Since each $J_{i}(i=1,2,...,6)$ is almost Hermitian structure\ with respect
to $g$, setting

\begin{equation}
\Phi _{i}(X,Y)=g(J_{i}X,Y),~\text{ }i=1,2,...,6,  \label{2.3}
\end{equation}

for any vector fields $X$ and $Y$, we see that $\Phi _{i}$ are 6 local
2-forms.

If the Levi-Civita connection $\nabla =\nabla ^{g}$ on $(M,g,V)$ preserves
the vector bundle $V$ by parallel transport, then $(M,g,V)$ is called a
Cliffordian K\"{a}hler manifold, and an almost Cliffordian structure $\Phi
_{i}$ of $M$ is called a Cliffordian K\"{a}hler structure. A Clifford K\"{a}%
hler manifold is Riemannian manifold ($M^{8n},g$)$.$ For example, we say
that $\mathbf{R}^{8n}$ is the simplest example of Clifford K\"{a}hler
manifold. Suppose that let $\left\{
x_{i},x_{n+i},x_{2n+i},x_{3n+i},x_{4n+i},x_{5n+i},x_{6n+i},x_{7n+i}\right\}
, $ $i=\overline{1,n}$ be a real coordinate system on $\mathbf{R}^{8n}.$
Then we define by $\left\{ \frac{\partial }{\partial x_{i}},\frac{\partial }{%
\partial x_{n+i}},\frac{\partial }{\partial x_{2n+i}},\frac{\partial }{%
\partial x_{3n+i}},\frac{\partial }{\partial x_{4n+i}},\frac{\partial }{%
\partial x_{5n+i}},\frac{\partial }{\partial x_{6n+i}},\frac{\partial }{%
\partial x_{7n+i}}\right\} $ and $%
\{dx_{i},dx_{n+i},dx_{2n+i},dx_{3n+i},dx_{4n+i},dx_{5n+i},dx_{6n+i},dx_{7n+i}\}
$ be natural bases over $\mathbf{R}$ of the tangent space $T(\mathbf{R}%
^{8n}) $ and the cotangent space $T^{\ast }(\mathbf{R}^{8n})$ of $\mathbf{R}%
^{8n},$ respectively$.$ By structures $J_{1},J_{2},J_{3}$, the following
expressions are obtained%
\begin{equation}
\begin{array}{c}
J_{1}(\frac{\partial }{\partial x_{i}})=\frac{\partial }{\partial x_{n+i}},%
\text{ }J_{1}(\frac{\partial }{\partial x_{n+i}})=-\frac{\partial }{\partial
x_{i}},\text{ }J_{1}(\frac{\partial }{\partial x_{2n+i}})=\frac{\partial }{%
\partial x_{4n+i}},\text{ }J_{1}(\frac{\partial }{\partial x_{3n+i}})=\frac{%
\partial }{\partial x_{5n+i}}, \\
J_{1}(\frac{\partial }{\partial x_{4n+i}})=-\frac{\partial }{\partial
x_{2n+i}},\text{ }J_{1}(\frac{\partial }{\partial x_{5n+i}})=-\frac{\partial
}{\partial x_{3n+i}},\text{ }J_{1}(\frac{\partial }{\partial x_{6n+i}})=%
\frac{\partial }{\partial x_{7n+i}},\text{ }J_{1}(\frac{\partial }{\partial
x_{7n+i}})=-\frac{\partial }{\partial x_{6n+i}}, \\
J_{2}(\frac{\partial }{\partial x_{i}})=\frac{\partial }{\partial x_{2n+i}},%
\text{ }J_{2}(\frac{\partial }{\partial x_{n+i}})=-\frac{\partial }{\partial
x_{4n+i}},\text{ }J_{2}(\frac{\partial }{\partial x_{2n+i}})=-\frac{\partial
}{\partial x_{i}},\text{ }J_{2}(\frac{\partial }{\partial x_{3n+i}})=\frac{%
\partial }{\partial x_{6n+i}}, \\
J_{2}(\frac{\partial }{\partial x_{4n+i}})=\frac{\partial }{\partial x_{n+i}}%
,\text{ }J_{2}(\frac{\partial }{\partial x_{5n+i}})=-\frac{\partial }{%
\partial x_{7n+i}},\text{ }J_{2}(\frac{\partial }{\partial x_{6n+i}})=-\frac{%
\partial }{\partial x_{3n+i}},\text{ }J_{2}(\frac{\partial }{\partial
x_{7n+i}})=\frac{\partial }{\partial x_{5n+i}}, \\
J_{3}(\frac{\partial }{\partial x_{i}})=\frac{\partial }{\partial x_{3n+i}},%
\text{ }J_{3}(\frac{\partial }{\partial x_{n+i}})=-\frac{\partial }{\partial
x_{5n+i}},\text{ }J_{3}(\frac{\partial }{\partial x_{2n+i}})=-\frac{\partial
}{\partial x_{6n+i}},\text{ }J_{3}(\frac{\partial }{\partial x_{3n+i}})=-%
\frac{\partial }{\partial x_{i}}, \\
J_{3}(\frac{\partial }{\partial x_{4n+i}})=\frac{\partial }{\partial x_{7n+i}%
},\text{ }J_{3}(\frac{\partial }{\partial x_{5n+i}})=\frac{\partial }{%
\partial x_{n+i}},\text{ }J_{3}(\frac{\partial }{\partial x_{6n+i}})=\frac{%
\partial }{\partial x_{2n+i}},\text{ }J_{3}(\frac{\partial }{\partial
x_{7n+i}})=-\frac{\partial }{\partial x_{4n+i}}.%
\end{array}
\label{2.4}
\end{equation}

\section{Lagrangian Mechanics}

In this section, we obtain Euler-Lagrange equations for quantum and
classical mechanics by means of a canonical local basis $\{J_{1},J_{2},J_{3}%
\}$ of $V$ on the standard Cliffordian K\"{a}hler manifold $(\mathbf{R}%
^{8n},V).$

Firstly, let $J_{1}$ take a local basis component on the standard
Cliffordian K\"{a}hler manifold $(\mathbf{R}^{8n},V),$ and $\left\{
x_{i},x_{n+i},x_{2n+i},x_{3n+i},x_{4n+i},x_{5n+i},x_{6n+i},x_{7n+i}\right\}
, $ $i=\overline{1,n}$ be its coordinate functions. Let semispray be the
vector field $\xi $ determined by%
\begin{equation}
\begin{array}{c}
\xi =X^{i}\frac{\partial }{\partial x_{i}}+X^{n+i}\frac{\partial }{\partial
x_{n+i}}+X^{2n+i}\frac{\partial }{\partial x_{2n+i}}+X^{3n+i}\frac{\partial
}{\partial x_{3n+i}} \\
+X^{4n+i}\frac{\partial }{\partial x_{4n+i}}+X^{5n+i}\frac{\partial }{%
\partial x_{5n+i}}+X^{6n+i}\frac{\partial }{\partial x_{6n+i}}+X^{7n+i}\frac{%
\partial }{\partial x_{7n+i}},%
\end{array}
\label{3.1}
\end{equation}%
where%
\begin{eqnarray*}
X^{i} &=&\overset{.}{x_{i}},X^{n+i}=\overset{.}{x}_{n+i},X^{2n+i}=\overset{.}%
{x}_{2n+i},X^{3n+i}=\overset{.}{x}_{3n+i}, \\
X^{4n+i} &=&\overset{.}{x_{4n+i}},X^{5n+i}=\overset{.}{x}_{5n+i},X^{6n+i}=%
\overset{.}{x}_{6n+i},X^{7n+i}=\overset{.}{x}_{7n+i}
\end{eqnarray*}

and the dot indicates the derivative with respect to time $t$. The vector
fields defined by%
\begin{equation}
\begin{array}{c}
V_{J_{1}}=J_{1}(\xi )=X^{i}\frac{\partial }{\partial x_{n+i}}-X^{n+i}\frac{%
\partial }{\partial x_{i}}+X^{2n+i}\frac{\partial }{\partial x_{4n+i}}%
+X^{3n+i}\frac{\partial }{\partial x_{5n+i}} \\
-X^{4n+i}\frac{\partial }{\partial x_{2n+i}}-X^{5n+i}\frac{\partial }{%
\partial x_{3n+i}}+X^{6n+i}\frac{\partial }{\partial x_{7n+i}}-X^{7n+i}\frac{%
\partial }{\partial x_{6n+i}},%
\end{array}
\label{3.2}
\end{equation}%
is called \textit{Liouville vector field} on the standard Cliffordian K\"{a}%
hler manifold $(\mathbf{R}^{8n},V)$. The maps given by $T,P:\mathbf{R}%
^{8n}\rightarrow \mathbf{R}$ such that%
\begin{equation*}
T=\frac{1}{2}m_{i}(\overset{.}{x_{i}}^{2}+\overset{.}{x}%
_{n+i}^{2}+x_{2n+i}^{2}+\overset{.}{x}_{3n+i}^{2}+\overset{.}{x_{4n+i}}^{2}+%
\overset{.}{x}_{5n+i}^{2}+x_{6n+i}^{2}+\overset{.}{x}_{7n+i}^{2}),\text{ \ }%
P=m_{i}gh
\end{equation*}

are called \textit{the kinetic energy} and \textit{the potential energy of
the system,} respectively.\textit{\ }Here\textit{\ }$m_{i},g$ and $h$ stand
for mass of a mechanical system having $m$ particles, the gravity
acceleration and distance to the origin of a mechanical system on the
standard Cliffordian K\"{a}hler manifold $(\mathbf{R}^{8n},V)$,
respectively. Then $L:\mathbf{R}^{8n}\rightarrow R$ is a map that satisfies
the conditions; i) $L=T-P$ is a \textit{Lagrangian function, ii)} the
function given by $E_{L}^{J_{1}}=V_{J_{1}}(L)-L,$ is\textit{\ energy function%
}.

The operator $i_{J_{1}}$ induced by $J_{1}$ and given by%
\begin{equation}
i_{J_{1}}\omega (X_{1},X_{2},...,X_{r})=\sum_{i=1}^{r}\omega
(X_{1},...,J_{1}X_{i},...,X_{r}),  \label{3.3}
\end{equation}

is said to be \textit{vertical derivation, }where $\omega \in \wedge ^{r}%
\mathbf{R}^{8n},$ $X_{i}\in \chi (\mathbf{R}^{8n}).$ The \textit{vertical
differentiation} $d_{J_{1}}$ is defined by%
\begin{equation}
d_{J_{1}}=[i_{J_{1}},d]=i_{J_{1}}d-di_{J_{1}}  \label{3.4}
\end{equation}%
where $d$ is the usual exterior derivation. For $J_{1}$ , the closed
Cliffordian K\"{a}hler form is the closed 2-form given by $\Phi
_{L}^{J_{1}}=-dd_{_{J_{1}}}L$ such that%
\begin{eqnarray*}
d_{_{J_{1}}} &=&\frac{\partial }{\partial x_{n+i}}dx_{i}-\frac{\partial }{%
\partial x_{i}}dx_{n+i}+\frac{\partial }{\partial x_{4n+i}}dx_{2n+i}+\frac{%
\partial }{\partial x_{5n+i}}dx_{3n+i} \\
&&-\frac{\partial }{\partial x_{2n+i}}dx_{4n+i}-\frac{\partial }{\partial
x_{3n+i}}dx_{5n+i}+\frac{\partial }{\partial x_{7n+i}}dx_{6n+i}-\frac{%
\partial }{\partial x_{6n+i}}dx_{7n+i}
\end{eqnarray*}

defined by operator%
\begin{equation}
d_{_{J_{1}}}:\mathcal{F}(\mathbf{R}^{8n})\rightarrow \wedge ^{1}{}\mathbf{R}%
^{8n}.  \label{3.5}
\end{equation}

Then

$\Phi _{L}^{J_{1}}=-\frac{\partial ^{2}L}{\partial x_{j}\partial x_{n+i}}%
dx_{j}\wedge dx_{i}+\frac{\partial ^{2}L}{\partial x_{j}\partial x_{i}}%
dx_{j}\wedge dx_{n+i}-\frac{\partial ^{2}L}{\partial x_{j}\partial x_{4n+i}}%
dx_{j}\wedge dx_{2n+i}$

$-\frac{\partial ^{2}L}{\partial x_{j}\partial x_{5n+i}}dx_{j}\wedge
dx_{3n+i}+\frac{\partial ^{2}L}{\partial x_{j}\partial x_{2n+i}}dx_{j}\wedge
dx_{4n+i}+\frac{\partial ^{2}L}{\partial x_{j}\partial x_{3n+i}}dx_{j}\wedge
dx_{5n+i}$

$-\frac{\partial ^{2}L}{\partial x_{j}\partial x_{7n+i}}dx_{j}\wedge
dx_{6n+i}+\frac{\partial ^{2}L}{\partial x_{j}\partial x_{6n+i}}dx_{j}\wedge
dx_{7n+i}-\frac{\partial ^{2}L}{\partial x_{n+j}\partial x_{n+i}}%
dx_{n+j}\wedge dx_{i}$

$+\frac{\partial ^{2}L}{\partial x_{n+j}\partial x_{i}}dx_{n+j}\wedge
dx_{n+i}-\frac{\partial ^{2}L}{\partial x_{n+j}\partial x_{4n+i}}%
dx_{n+j}\wedge dx_{2n+i}-\frac{\partial ^{2}L}{\partial x_{n+j}\partial
x_{5n+i}}dx_{n+j}\wedge dx_{3n+i}$

$+\frac{\partial ^{2}L}{\partial x_{n+j}\partial x_{2n+i}}dx_{n+j}\wedge
dx_{4n+i}+\frac{\partial ^{2}L}{\partial x_{n+j}\partial x_{3n+i}}%
dx_{n+j}\wedge dx_{5n+i}-\frac{\partial ^{2}L}{\partial x_{n+j}\partial
x_{7n+i}}dx_{n+j}\wedge dx_{6n+i}$

$+\frac{\partial ^{2}L}{\partial x_{n+j}\partial x_{6n+i}}dx_{n+j}\wedge
dx_{7n+i}-\ \frac{\partial ^{2}L}{\partial x_{2n+j}\partial x_{n+i}}%
dx_{2n+j}\wedge dx_{i}+\frac{\partial ^{2}L}{\partial x_{2n+j}\partial x_{i}}%
dx_{2n+j}\wedge dx_{n+i}$

$-\frac{\partial ^{2}L}{\partial x_{2n+j}\partial x_{4n+i}}dx_{2n+j}\wedge
dx_{2n+i}-\frac{\partial ^{2}L}{\partial x_{2n+j}\partial x_{5n+i}}%
dx_{2n+j}\wedge dx_{3n+i}+\frac{\partial ^{2}L}{\partial x_{2n+j}\partial
x_{2n+i}}dx_{2n+j}\wedge dx_{4n+i}$

$+\frac{\partial ^{2}L}{\partial x_{2n+j}\partial x_{3n+i}}dx_{2n+j}\wedge
dx_{5n+i}-\frac{\partial ^{2}L}{\partial x_{2n+j}\partial x_{7n+i}}%
dx_{2n+j}\wedge dx_{6n+i}+\frac{\partial ^{2}L}{\partial x_{2n+j}\partial
x_{6n+i}}dx_{2n+j}\wedge dx_{7n+i}$

$-\frac{\partial ^{2}L}{\partial x_{3n+j}\partial x_{n+i}}dx_{3n+j}\wedge
dx_{i}+\frac{\partial ^{2}L}{\partial x_{3n+j}\partial x_{i}}dx_{3n+j}\wedge
dx_{n+i}-\frac{\partial ^{2}L}{\partial x_{3n+j}\partial x_{4n+i}}%
dx_{3n+j}\wedge dx_{2n+i}$

$-\frac{\partial ^{2}L}{\partial x_{3n+j}\partial x_{5n+i}}dx_{3n+j}\wedge
dx_{3n+i}+\frac{\partial ^{2}L}{\partial x_{3n+j}\partial x_{2n+i}}%
dx_{3n+j}\wedge dx_{4n+i}+\frac{\partial ^{2}L}{\partial x_{3n+j}\partial
x_{3n+i}}dx_{3n+j}\wedge dx_{5n+i}$

$-\frac{\partial ^{2}L}{\partial x_{3n+j}\partial x_{7n+i}}dx_{3n+j}\wedge
dx_{6n+i}+\frac{\partial ^{2}L}{\partial x_{3n+j}\partial x_{6n+i}}%
dx_{3n+j}\wedge dx_{7n+i}-\frac{\partial L}{\partial x_{4n+j}\partial x_{n+i}%
}dx_{4n+j}\wedge dx_{i}$

$+\frac{\partial L}{\partial x_{4n+j}\partial x_{i}}dx_{4n+j}\wedge dx_{n+i}-%
\frac{\partial L}{\partial x_{4n+j}\partial x_{4n+i}}dx_{4n+j}\wedge
dx_{2n+i}-\frac{\partial L}{\partial x_{4n+j}\partial x_{5n+i}}%
dx_{4n+j}\wedge dx_{3n+i}$

$+\frac{\partial ^{2}L}{\partial x_{4n+j}\partial x_{2n+i}}dx_{4n+j}\wedge
dx_{4n+i}+\frac{\partial ^{2}L}{\partial x_{4n+j}\partial x_{3n+i}}%
dx_{4n+j}\wedge dx_{5n+i}-\frac{\partial ^{2}L}{\partial x_{4n+j}\partial
x_{7n+i}}dx_{4n+j}\wedge dx_{6n+i}$

$+\frac{\partial ^{2}L}{\partial x_{4n+j}\partial x_{6n+i}}dx_{4n+j}\wedge
dx_{7n+i}-\frac{\partial ^{2}L}{\partial x_{5n+j}\partial x_{n+i}}%
dx_{5n+j}\wedge dx_{i}+\frac{\partial ^{2}L}{\partial x_{5n+j}\partial x_{i}}%
dx_{5n+j}\wedge dx_{n+i}$

$-\frac{\partial ^{2}L}{\partial x_{5n+j}\partial x_{4n+i}}dx_{5n+j}\wedge
dx_{2n+i}-\frac{\partial ^{2}L}{\partial x_{5n+j}\partial x_{5n+i}}%
dx_{5n+j}\wedge dx_{3n+i}+\frac{\partial ^{2}L}{\partial x_{5n+j}\partial
x_{2n+i}}dx_{5n+j}\wedge dx_{4n+i}$

$+\frac{\partial ^{2}L}{\partial x_{5n+j}\partial x_{3n+i}}dx_{5n+j}\wedge
dx_{5n+i}-\frac{\partial ^{2}L}{\partial x_{5n+j}\partial x_{7n+i}}%
dx_{5n+j}\wedge dx_{6n+i}+\frac{\partial ^{2}L}{\partial x_{5n+j}\partial
x_{6n+i}}dx_{5n+j}\wedge dx_{7n+i}$

$-\frac{\partial ^{2}L}{\partial x_{6n+j}\partial x_{n+i}}dx_{6n+j}\wedge
dx_{i}+\frac{\partial ^{2}L}{\partial x_{6n+j}\partial x_{i}}dx_{6n+j}\wedge
dx_{n+i}-\frac{\partial ^{2}L}{\partial x_{6n+j}\partial x_{4n+i}}%
dx_{6n+j}\wedge dx_{2n+i}$

$-\frac{\partial ^{2}L}{\partial x_{6n+j}\partial x_{5n+i}}dx_{6n+j}\wedge
dx_{3n+i}+\frac{\partial ^{2}L}{\partial x_{6n+j}\partial x_{2n+i}}%
dx_{6n+j}\wedge dx_{4n+i}+\frac{\partial ^{2}L}{\partial x_{6n+j}\partial
x_{3n+i}}dx_{6n+j}\wedge dx_{5n+i}$

$-\frac{\partial ^{2}L}{\partial x_{6n+j}\partial x_{7n+i}}dx_{6n+j}\wedge
dx_{6n+i}+\frac{\partial ^{2}L}{\partial x_{6n+j}\partial x_{6n+i}}%
dx_{6n+j}\wedge dx_{7n+i}-\frac{\partial ^{2}L}{\partial x_{7n+j}\partial
x_{n+i}}dx_{7n+j}\wedge dx_{i}$

$+\frac{\partial ^{2}L}{\partial x_{7n+j}\partial x_{i}}dx_{7n+j}\wedge
dx_{n+i}-\frac{\partial ^{2}L}{\partial x_{7n+j}\partial x_{4n+i}}%
dx_{7n+j}\wedge dx_{2n+i}-\frac{\partial ^{2}L}{\partial x_{7n+j}\partial
x_{5n+i}}dx_{7n+j}\wedge dx_{3n+i}$

$+\frac{\partial ^{2}L}{\partial x_{7n+j}\partial x_{2n+i}}dx_{7n+j}\wedge
dx_{4n+i}+\frac{\partial ^{2}L}{\partial x_{7n+j}\partial x_{3n+i}}%
dx_{7n+j}\wedge dx_{5n+i}-\frac{\partial ^{2}L}{\partial x_{7n+j}\partial
x_{7n+i}}dx_{7n+j}\wedge dx_{6n+i}$

$+\frac{\partial ^{2}L}{\partial x_{7n+j}\partial x_{6n+i}}dx_{7n+j}\wedge
dx_{7n+i}$

Let $\xi $ be the second order differential equation by given \textbf{Eq. }(%
\ref{1.1}) and defined by \textbf{Eq. }(\ref{3.1}) and

$i_{\xi }\Phi _{L}^{J_{1}}=-X^{i}\frac{\partial ^{2}L}{\partial
x_{j}\partial x_{n+i}}\delta _{i}^{j}dx_{i}+X^{i}\frac{\partial ^{2}L}{%
\partial x_{j}\partial x_{n+i}}dx_{j}+X^{i}\frac{\partial ^{2}L}{\partial
x_{j}\partial x_{i}}\delta _{i}^{j}dx_{n+i}-X^{n+i}\frac{\partial ^{2}L}{%
\partial x_{j}\partial x_{i}}dx_{j}$

$-X^{i}\frac{\partial ^{2}L}{\partial x_{j}\partial x_{4n+i}}\delta
_{i}^{j}dx_{2n+i}+X^{2n+i}\frac{\partial ^{2}L}{\partial x_{j}\partial
x_{4n+i}}dx_{j}-X^{i}\frac{\partial ^{2}L}{\partial x_{j}\partial x_{5n+i}}%
\delta _{i}^{j}dx_{3n+i}+X^{3n+i}\frac{\partial ^{2}L}{\partial
x_{j}\partial x_{5n+i}}dx_{j}$

$+X^{i}\frac{\partial ^{2}L}{\partial x_{j}\partial x_{2n+i}}\delta
_{i}^{j}dx_{4n+i}-X^{4n+i}\frac{\partial ^{2}L}{\partial x_{j}\partial
x_{2n+i}}dx_{j}+X^{i}\frac{\partial ^{2}L}{\partial x_{j}\partial x_{3n+i}}%
\delta _{i}^{j}dx_{5n+i}-X^{5n+i}\frac{\partial ^{2}L}{\partial
x_{j}\partial x_{3n+i}}dx_{j}$

$-X^{i}\frac{\partial ^{2}L}{\partial x_{j}\partial x_{7n+i}}\delta
_{i}^{j}dx_{6n+i}+X^{6n+i}\frac{\partial ^{2}L}{\partial x_{j}\partial
x_{7n+i}}dx_{j}+X^{i}\frac{\partial ^{2}L}{\partial x_{j}\partial x_{6n+i}}%
\delta _{i}^{j}dx_{7n+i}-X^{7n+i}\frac{\partial ^{2}L}{\partial
x_{j}\partial x_{6n+i}}dx_{j}$

$-X^{n+i}\frac{\partial ^{2}L}{\partial x_{n+j}\partial x_{n+i}}\delta
_{n+i}^{n+j}dx_{i}+X^{n+i}\frac{\partial ^{2}L}{\partial x_{n+j}\partial
x_{n+i}}dx_{n+j}+X^{n+i}\frac{\partial ^{2}L}{\partial x_{n+j}\partial x_{i}}%
\delta _{n+i}^{n+j}dx_{n+i}-X^{n+i}\frac{\partial ^{2}L}{\partial
x_{n+j}\partial x_{i}}dx_{n+j}$

$-X^{n+i}\frac{\partial ^{2}L}{\partial x_{n+j}\partial x_{4n+i}}\delta
_{n+i}^{n+j}dx_{2n+i}\ +X^{2n+i}\frac{\partial ^{2}L}{\partial
x_{n+j}\partial x_{4n+i}}dx_{n+j}-X^{n+i}\frac{\partial ^{2}L}{\partial
x_{n+j}\partial x_{5n+i}}\delta _{n+i}^{n+j}dx_{3n+i}$

$+X^{3n+i}\frac{\partial ^{2}L}{\partial x_{n+j}\partial x_{5n+i}}%
dx_{n+j}+X^{n+i}\frac{\partial ^{2}L}{\partial x_{n+j}\partial x_{2n+i}}%
\delta _{n+i}^{n+j}dx_{4n+i}-X^{4n+i}\frac{\partial ^{2}L}{\partial
x_{n+j}\partial x_{2n+i}}dx_{n+j}$

$+X^{n+i}\frac{\partial ^{2}L}{\partial x_{n+j}\partial x_{3n+i}}\delta
_{n+i}^{n+j}dx_{5n+i}-X^{5n+i}\frac{\partial ^{2}L}{\partial x_{n+j}\partial
x_{3n+i}}dx_{n+j}-X^{n+i}\frac{\partial ^{2}L}{\partial x_{n+j}\partial
x_{7n+i}}\delta _{n+i}^{n+j}dx_{6n+i}$

$+X^{6n+i}\frac{\partial ^{2}L}{\partial x_{n+j}\partial x_{7n+i}}%
dx_{n+j}+X^{n+i}\frac{\partial ^{2}L}{\partial x_{n+j}\partial x_{6n+i}}%
\delta _{n+i}^{n+j}dx_{7n+i}-X^{7n+i}\frac{\partial ^{2}L}{\partial
x_{n+j}\partial x_{6n+i}}dx_{n+j}$

$-\ X^{2n+i}\frac{\partial ^{2}L}{\partial x_{2n+j}\partial x_{n+i}}\delta
_{2n+i}^{2n+j}dx_{i}+\ X^{i}\frac{\partial ^{2}L}{\partial x_{2n+j}\partial
x_{n+i}}dx_{2n+j}+X^{2n+i}\frac{\partial ^{2}L}{\partial x_{2n+j}\partial
x_{i}}\delta _{2n+i}^{2n+j}dx_{n+i}$

$-X^{n+i}\frac{\partial ^{2}L}{\partial x_{2n+j}\partial x_{i}}%
dx_{2n+j}-X^{2n+i}\frac{\partial ^{2}L}{\partial x_{2n+j}\partial x_{4n+i}}%
\delta _{2n+i}^{2n+j}dx_{2n+i}+X^{2n+i}\frac{\partial ^{2}L}{\partial
x_{2n+j}\partial x_{4n+i}}dx_{2n+j}$

$-X^{2n+i}\frac{\partial ^{2}L}{\partial x_{2n+j}\partial x_{5n+i}}\delta
_{2n+i}^{2n+j}dx_{3n+i}+X^{3n+i}\frac{\partial ^{2}L}{\partial
x_{2n+j}\partial x_{5n+i}}dx_{2n+j}+X^{2n+i}\frac{\partial ^{2}L}{\partial
x_{2n+j}\partial x_{2n+i}}\delta _{2n+i}^{2n+j}dx_{4n+i}$

$-X^{4n+i}\frac{\partial ^{2}L}{\partial x_{2n+j}\partial x_{2n+i}}%
dx_{2n+j}+X^{2n+i}\frac{\partial ^{2}L}{\partial x_{2n+j}\partial x_{3n+i}}%
\delta _{2n+i}^{2n+j}dx_{5n+i}-X^{5n+i}\frac{\partial ^{2}L}{\partial
x_{2n+j}\partial x_{3n+i}}dx_{2n+j}$

$-X^{2n+i}\frac{\partial ^{2}L}{\partial x_{2n+j}\partial x_{7n+i}}\delta
_{2n+i}^{2n+j}dx_{6n+i}+X^{6n+i}\frac{\partial ^{2}L}{\partial
x_{2n+j}\partial x_{7n+i}}dx_{2n+j}+X^{2n+i}\frac{\partial ^{2}L}{\partial
x_{2n+j}\partial x_{6n+i}}\delta _{2n+i}^{2n+j}dx_{7n+i}$

$-X^{7n+i}\frac{\partial ^{2}L}{\partial x_{2n+j}\partial x_{6n+i}}%
dx_{2n+j}-X^{3n+i}\frac{\partial ^{2}L}{\partial x_{3n+j}\partial x_{n+i}}%
\delta _{3n+i}^{3n+j}dx_{i}+X^{i}\frac{\partial ^{2}L}{\partial
x_{3n+j}\partial x_{n+i}}dx_{i}$

$+X^{3n+i}\frac{\partial ^{2}L}{\partial x_{3n+j}\partial x_{i}}\delta
_{3n+i}^{3n+j}dx_{n+i}-X^{n+i}\frac{\partial ^{2}L}{\partial
x_{3n+j}\partial x_{i}}dx_{3n+j}-X^{3n+i}\frac{\partial ^{2}L}{\partial
x_{3n+j}\partial x_{4n+i}}\delta _{3n+i}^{3n+j}dx_{2n+i}$

$+X^{2n+i}\frac{\partial ^{2}L}{\partial x_{3n+j}\partial x_{4n+i}}%
dx_{3n+j}-X^{3n+i}\frac{\partial ^{2}L}{\partial x_{3n+j}\partial x_{5n+i}}%
\delta _{3n+i}^{3n+j}dx_{3n+i}+X^{3n+i}\frac{\partial ^{2}L}{\partial
x_{3n+j}\partial x_{5n+i}}dx_{3n+j}$

$+X^{3n+i}\frac{\partial ^{2}L}{\partial x_{3n+j}\partial x_{2n+i}}\delta
_{3n+i}^{3n+j}dx_{4n+i}-X^{4n+i}\frac{\partial ^{2}L}{\partial
x_{3n+j}\partial x_{2n+i}}dx_{3n+j}+X^{3n+i}\frac{\partial ^{2}L}{\partial
x_{3n+j}\partial x_{3n+i}}\delta _{3n+i}^{3n+j}dx_{5n+i}$

$-X^{5n+i}\frac{\partial ^{2}L}{\partial x_{3n+j}\partial x_{3n+i}}%
dx_{3n+j}-X^{3n+i}\frac{\partial ^{2}L}{\partial x_{3n+j}\partial x_{7n+i}}%
\delta _{3n+i}^{3n+j}dx_{6n+i}+X^{6n+i}\frac{\partial ^{2}L}{\partial
x_{3n+j}\partial x_{7n+i}}dx_{3n+j}$

$+X^{3n+i}\frac{\partial ^{2}L}{\partial x_{3n+j}\partial x_{6n+i}}\delta
_{3n+i}^{3n+j}dx_{7n+i}-X^{7n+i}\frac{\partial ^{2}L}{\partial
x_{3n+j}\partial x_{6n+i}}dx_{3n+j}-X^{4n+i}\frac{\partial L}{\partial
x_{4n+j}\partial x_{n+i}}\delta _{4n+i}^{4n+j}dx_{i}$

$\ +X^{i}\frac{\partial L}{\partial x_{4n+j}\partial x_{n+i}}%
dx_{4n+j}+X^{4n+i}\frac{\partial L}{\partial x_{4n+j}\partial x_{i}}\delta
_{4n+i}^{4n+j}dx_{n+i}-X^{n+i}\frac{\partial L}{\partial x_{4n+j}\partial
x_{i}}dx_{4n+j}$

$-X^{4n+i}\frac{\partial L}{\partial x_{4n+j}\partial x_{4n+i}}\delta
_{4n+i}^{4n+j}dx_{2n+i}\ \ +X^{2n+i}\frac{\partial L}{\partial
x_{4n+j}\partial x_{4n+i}}dx_{4n+j}-X^{4n+i}\frac{\partial L}{\partial
x_{4n+j}\partial x_{5n+i}}\delta _{4n+i}^{4n+j}dx_{3n+i}$

$+X^{3n+i}\frac{\partial L}{\partial x_{4n+j}\partial x_{5n+i}}%
dx_{4n+j}+X^{4n+i}\frac{\partial ^{2}L}{\partial x_{4n+j}\partial x_{2n+i}}%
\delta _{4n+i}^{4n+j}dx_{4n+i}-X^{4n+i}\frac{\partial ^{2}L}{\partial
x_{4n+j}\partial x_{2n+i}}dx_{4n+j}$

$+X^{4n+i}\frac{\partial ^{2}L}{\partial x_{4n+j}\partial x_{3n+i}}\delta
_{4n+i}^{4n+j}dx_{5n+i}-X^{5n+i}\frac{\partial ^{2}L}{\partial
x_{4n+j}\partial x_{3n+i}}dx_{4n+j}-X^{4n+i}\frac{\partial ^{2}L}{\partial
x_{4n+j}\partial x_{7n+i}}\delta _{4n+i}^{4n+j}dx_{6n+i}$

$+X^{6n+i}\frac{\partial ^{2}L}{\partial x_{4n+j}\partial x_{7n+i}}dx_{4n+j}+%
\frac{\partial ^{2}L}{\partial x_{4n+j}\partial x_{6n+i}}\delta
_{4n+i}^{4n+j}dx_{7n+i}-X^{7n+i}\frac{\partial ^{2}L}{\partial
x_{4n+j}\partial x_{6n+i}}dx_{4n+j}$

$-X^{5n+i}\frac{\partial ^{2}L}{\partial x_{5n+j}\partial x_{n+i}}\delta
_{5n+i}^{5n+j}dx_{i}+X^{i}\frac{\partial ^{2}L}{\partial x_{5n+j}\partial
x_{n+i}}dx_{5n+j}+X^{5n+i}\frac{\partial ^{2}L}{\partial x_{5n+j}\partial
x_{i}}\delta _{5n+i}^{5n+j}dx_{n+i}$

$-X^{n+i}\frac{\partial ^{2}L}{\partial x_{5n+j}\partial x_{i}}%
dx_{5n+j}-X^{5n+i}\frac{\partial ^{2}L}{\partial x_{5n+j}\partial x_{4n+i}}%
\delta _{5n+i}^{5n+j}dx_{2n+i}+X^{2n+i}\frac{\partial ^{2}L}{\partial
x_{5n+j}\partial x_{4n+i}}dx_{5n+j}$

$-X^{5n+i}\frac{\partial ^{2}L}{\partial x_{5n+j}\partial x_{5n+i}}\delta
_{5n+i}^{5n+j}dx_{3n+i}+X^{3n+i}\frac{\partial ^{2}L}{\partial
x_{5n+j}\partial x_{5n+i}}dx_{5n+j}+X^{5n+i}\frac{\partial ^{2}L}{\partial
x_{5n+j}\partial x_{2n+i}}\delta _{5n+i}^{5n+j}dx_{4n+i}$

$-X^{4n+i}\frac{\partial ^{2}L}{\partial x_{5n+j}\partial x_{2n+i}}%
dx_{5n+j}+X^{5n+i}\frac{\partial ^{2}L}{\partial x_{5n+j}\partial x_{3n+i}}%
\delta _{5n+i}^{5n+j}dx_{5n+i}-X^{5n+i}\frac{\partial ^{2}L}{\partial
x_{5n+j}\partial x_{3n+i}}dx_{5n+j}$

$-X^{5n+i}\frac{\partial ^{2}L}{\partial x_{5n+j}\partial x_{7n+i}}\delta
_{5n+i}^{5n+j}dx_{6n+i}+X^{6n+i}\frac{\partial ^{2}L}{\partial
x_{5n+j}\partial x_{7n+i}}dx_{5n+j}+X^{5n+i}\frac{\partial ^{2}L}{\partial
x_{5n+j}\partial x_{6n+i}}\delta _{5n+i}^{5n+j}dx_{7n+i}$

$-X^{7n+i}\frac{\partial ^{2}L}{\partial x_{5n+j}\partial x_{6n+i}}%
dx_{5n+j}-X^{6n+i}\frac{\partial ^{2}L}{\partial x_{6n+j}\partial x_{n+i}}%
\delta _{6n+i}^{6n+j}dx_{i}+X^{i}\frac{\partial ^{2}L}{\partial
x_{6n+j}\partial x_{n+i}}dx_{6n+j}$

$+X^{6n+i}\frac{\partial ^{2}L}{\partial x_{6n+j}\partial x_{i}}\delta
_{6n+i}^{6n+j}dx_{n+i}-X^{n+i}\frac{\partial ^{2}L}{\partial
x_{6n+j}\partial x_{i}}dx_{6n+j}-X^{6n+i}\frac{\partial ^{2}L}{\partial
x_{6n+j}\partial x_{4n+i}}\delta _{6n+i}^{6n+j}dx_{2n+i}$

$+X^{2n+i}\frac{\partial ^{2}L}{\partial x_{6n+j}\partial x_{4n+i}}%
dx_{6n+j}-X^{6n+i}\frac{\partial ^{2}L}{\partial x_{6n+j}\partial x_{5n+i}}%
\delta _{6n+i}^{6n+j}dx_{3n+i}+X^{3n+i}\frac{\partial ^{2}L}{\partial
x_{6n+j}\partial x_{5n+i}}dx_{6n+j}$

$+X^{6n+i}\frac{\partial ^{2}L}{\partial x_{6n+j}\partial x_{2n+i}}\delta
_{6n+i}^{6n+j}dx_{4n+i}-X^{4n+i}\frac{\partial ^{2}L}{\partial
x_{6n+j}\partial x_{2n+i}}dx_{6n+j}+X^{6n+i}\frac{\partial ^{2}L}{\partial
x_{6n+j}\partial x_{3n+i}}\delta _{6n+i}^{6n+j}dx_{5n+i}$

$-X^{5n+i}\frac{\partial ^{2}L}{\partial x_{6n+j}\partial x_{3n+i}}%
dx_{6n+j}-X^{6n+i}\frac{\partial ^{2}L}{\partial x_{6n+j}\partial x_{7n+i}}%
\delta _{6n+i}^{6n+j}dx_{6n+i}+X^{6n+i}\frac{\partial ^{2}L}{\partial
x_{6n+j}\partial x_{7n+i}}dx_{6n+j}$

$+X^{6n+i}\frac{\partial ^{2}L}{\partial x_{6n+j}\partial x_{6n+i}}\delta
_{6n+i}^{6n+j}dx_{7n+i}-X^{7n+i}\frac{\partial ^{2}L}{\partial
x_{6n+j}\partial x_{6n+i}}dx_{6n+j}-X^{7n+i}\frac{\partial ^{2}L}{\partial
x_{7n+j}\partial x_{n+i}}\delta _{7n+i}^{7n+j}dx_{i}$

$+X^{i}\frac{\partial ^{2}L}{\partial x_{7n+j}\partial x_{n+i}}%
dx_{7n+j}+X^{7n+i}\frac{\partial ^{2}L}{\partial x_{7n+j}\partial x_{i}}%
\delta _{7n+i}^{7n+j}dx_{n+i}-X^{n+i}\frac{\partial ^{2}L}{\partial
x_{7n+j}\partial x_{i}}dx_{7n+j}$

$-X^{7n+i}\frac{\partial ^{2}L}{\partial x_{7n+j}\partial x_{4n+i}}\delta
_{7n+i}^{7n+j}dx_{2n+i}+X^{2n+i}\frac{\partial ^{2}L}{\partial
x_{7n+j}\partial x_{4n+i}}dx_{7n+j}-X^{7n+i}\frac{\partial ^{2}L}{\partial
x_{7n+j}\partial x_{5n+i}}\delta _{7n+i}^{7n+j}dx_{3n+i}$

$+X^{3n+i}\frac{\partial ^{2}L}{\partial x_{7n+j}\partial x_{5n+i}}%
dx_{7n+j}+X^{7n+i}\frac{\partial ^{2}L}{\partial x_{7n+j}\partial x_{2n+i}}%
\delta _{7n+i}^{7n+j}dx_{4n+i}\ \ -X^{4n+i}\frac{\partial ^{2}L}{\partial
x_{7n+j}\partial x_{2n+i}}dx_{7n+j}$

$+X^{7n+i}\frac{\partial ^{2}L}{\partial x_{7n+j}\partial x_{3n+i}}\delta
_{7n+i}^{7n+j}dx_{5n+i}-X^{5n+i}\frac{\partial ^{2}L}{\partial
x_{7n+j}\partial x_{3n+i}}dx_{7n+j}-X^{7n+i}\frac{\partial ^{2}L}{\partial
x_{7n+j}\partial x_{7n+i}}\delta _{7n+i}^{7n+j}dx_{6n+i}$

$+X^{6n+i}\frac{\partial ^{2}L}{\partial x_{7n+j}\partial x_{7n+i}}%
dx_{7n+j}+X^{7n+i}\frac{\partial ^{2}L}{\partial x_{7n+j}\partial x_{6n+i}}%
\delta _{7n+i}^{7n+j}dx_{7n+i}-X^{7n+i}\frac{\partial ^{2}L}{\partial
x_{7n+j}\partial x_{6n+i}}dx_{7n+j}$

Since the closed standard Cliffordian K\"{a}hler form $\Phi _{L}^{J_{1}}$ on
$(\mathbf{R}^{8},V)$ is the symplectic structure, it is found

$E_{L}^{J_{1}}=V_{J_{1}}(L)-L=X^{i}\frac{\partial L}{\partial x_{n+i}}%
-X^{n+i}\frac{\partial L}{\partial x_{i}}+X^{2n+i}\frac{\partial L}{\partial
x_{4n+i}}+X^{3n+i}\frac{\partial L}{\partial x_{5n+i}}$

$\ \ \ \ \ \ \ \ \ \ \ \ \ \ \ \ \ \ \ \ \ \ \ \ \ \ \ \ -X^{4n+i}\frac{%
\partial L}{\partial x_{2n+i}}-X^{5n+i}\frac{\partial L}{\partial x_{3n+i}}%
+X^{6n+i}\frac{\partial L}{\partial x_{7n+i}}-X^{7n+i}\frac{\partial L}{%
\partial x_{6n+i}}-L$

and hence

$dE_{L}^{J_{1}}=X^{i}\frac{\partial ^{2}L}{\partial x_{j}\partial x_{n+i}}%
dx_{j}-X^{n+i}\frac{\partial ^{2}L}{\partial x_{j}\partial x_{i}}%
dx_{j}+X^{2n+i}\frac{\partial ^{2}L}{\partial x_{j}\partial x_{4n+i}}dx_{j}$

$+X^{3n+i}\frac{\partial ^{2}L}{\partial x_{j}\partial x_{5n+i}}%
dx_{j}-X^{4n+i}\frac{\partial ^{2}L}{\partial x_{j}\partial x_{2n+i}}%
dx_{j}-X^{5n+i}\frac{\partial ^{2}L}{\partial x_{j}\partial x_{3n+i}}dx_{j}$

$+X^{6n+i}\frac{\partial ^{2}L}{\partial x_{j}\partial x_{7n+i}}%
dx_{j}-X^{7n+i}\frac{\partial ^{2}L}{\partial x_{j}\partial x_{6n+i}}%
dx_{j}+X^{i}\frac{\partial ^{2}L}{\partial x_{n+j}\partial x_{n+i}}dx_{n+j}$

$-X^{n+i}\frac{\partial ^{2}L}{\partial x_{n+j}\partial x_{i}}%
dx_{n+j}+X^{2n+i}\frac{\partial ^{2}L}{\partial x_{n+j}\partial x_{4n+i}}%
dx_{n+j}+X^{3n+i}\frac{\partial ^{2}L}{\partial x_{n+j}\partial x_{5n+i}}%
dx_{n+j}$

$-X^{4n+i}\frac{\partial ^{2}L}{\partial x_{n+j}\partial x_{2n+i}}%
dx_{n+j}-X^{5n+i}\frac{\partial ^{2}L}{\partial x_{n+j}\partial x_{3n+i}}%
dx_{n+j}+X^{6n+i}\frac{\partial ^{2}L}{\partial x_{n+j}\partial x_{7n+i}}%
dx_{n+j}$

$-X^{7n+i}\frac{\partial ^{2}L}{\partial x_{n+j}\partial x_{6n+i}}%
dx_{n+j}+X^{i}\frac{\partial ^{2}L}{\partial x_{2n+j}\partial x_{n+i}}%
dx_{2n+j}-X^{n+i}\frac{\partial ^{2}L}{\partial x_{2n+j}\partial x_{i}}%
dx_{2n+j}$

$+X^{2n+i}\frac{\partial ^{2}L}{\partial x_{2n+j}\partial x_{4n+i}}%
dx_{2n+j}+X^{3n+i}\frac{\partial ^{2}L}{\partial x_{2n+j}\partial x_{5n+i}}%
dx_{2n+j}-X^{4n+i}\frac{\partial ^{2}L}{\partial x_{2n+j}\partial x_{2n+i}}%
dx_{2n+j}$

$-X^{5n+i}\frac{\partial ^{2}L}{\partial x_{2n+j}\partial x_{3n+i}}%
dx_{2n+j}+X^{6n+i}\frac{\partial ^{2}L}{\partial x_{2n+j}\partial x_{7n+i}}%
dx_{2n+j}-X^{7n+i}\frac{\partial ^{2}L}{\partial x_{2n+j}\partial x_{6n+i}}%
dx_{2n+j}$

$+X^{i}\frac{\partial ^{2}L}{\partial x_{3n+j}\partial x_{n+i}}%
dx_{3n+j}-X^{n+i}\frac{\partial ^{2}L}{\partial x_{3n+j}\partial x_{i}}%
dx_{3n+j}+X^{2n+i}\frac{\partial ^{2}L}{\partial x_{3n+j}\partial x_{4n+i}}%
dx_{3n+j}$

$+X^{3n+i}\frac{\partial ^{2}L}{\partial x_{3n+j}\partial x_{5n+i}}%
dx_{3n+j}-X^{4n+i}\frac{\partial ^{2}L}{\partial x_{3n+j}\partial x_{2n+i}}%
dx_{3n+j}-X^{5n+i}\frac{\partial ^{2}L}{\partial x_{3n+j}\partial x_{3n+i}}%
dx_{3n+j}$

$+X^{6n+i}\frac{\partial ^{2}L}{\partial x_{3n+j}\partial x_{7n+i}}%
dx_{3n+j}-X^{7n+i}\frac{\partial ^{2}L}{\partial x_{3n+j}\partial x_{6n+i}}%
dx_{3n+j}+X^{i}\frac{\partial ^{2}L}{\partial x_{4n+j}\partial x_{n+i}}%
dx_{4n+j}$

$-X^{n+i}\frac{\partial ^{2}L}{\partial x_{4n+j}\partial x_{i}}%
dx_{4n+j}+X^{2n+i}\frac{\partial ^{2}L}{\partial x_{4n+j}\partial x_{4n+i}}%
dx_{4n+j}+X^{3n+i}\frac{\partial ^{2}L}{\partial x_{4n+j}\partial x_{5n+i}}%
dx_{4n+j}$

$-X^{4n+i}\frac{\partial ^{2}L}{\partial x_{4n+j}\partial x_{2n+i}}%
dx_{4n+j}-X^{5n+i}\frac{\partial ^{2}L}{\partial x_{4n+j}\partial x_{3n+i}}%
dx_{4n+j}+X^{6n+i}\frac{\partial ^{2}L}{\partial x_{4n+j}\partial x_{7n+i}}%
dx_{4n+j}$

$-X^{7n+i}\frac{\partial ^{2}L}{\partial x_{4n+j}\partial x_{6n+i}}%
dx_{4n+j}+X^{i}\frac{\partial ^{2}L}{\partial x_{5n+j}\partial x_{n+i}}%
dx_{5n+j}-X^{n+i}\frac{\partial ^{2}L}{\partial x_{5n+j}\partial x_{i}}%
dx_{5n+j}$

$+X^{2n+i}\frac{\partial ^{2}L}{\partial x_{5n+j}\partial x_{4n+i}}%
dx_{5n+j}+X^{3n+i}\frac{\partial ^{2}L}{\partial x_{5n+j}\partial x_{5n+i}}%
dx_{5n+j}-X^{4n+i}\frac{\partial ^{2}L}{\partial x_{5n+j}\partial x_{2n+i}}%
dx_{5n+j}$

$-X^{5n+i}\frac{\partial ^{2}L}{\partial x_{5n+j}\partial x_{3n+i}}%
dx_{5n+j}+X^{6n+i}\frac{\partial ^{2}L}{\partial x_{5n+j}\partial x_{7n+i}}%
dx_{5n+j}-X^{7n+i}\frac{\partial ^{2}L}{\partial x_{5n+j}\partial x_{6n+i}}%
dx_{5n+j}$

$+X^{i}\frac{\partial ^{2}L}{\partial x_{6n+j}\partial x_{n+i}}%
dx_{6n+j}-X^{n+i}\frac{\partial ^{2}L}{\partial x_{6n+j}\partial x_{i}}%
dx_{6n+j}+X^{2n+i}\frac{\partial ^{2}L}{\partial x_{6n+j}\partial x_{4n+i}}%
dx_{6n+j}$

$+X^{3n+i}\frac{\partial ^{2}L}{\partial x_{6n+j}\partial x_{5n+i}}%
dx_{6n+j}-X^{4n+i}\frac{\partial ^{2}L}{\partial x_{6n+j}\partial x_{2n+i}}%
dx_{6n+j}-X^{5n+i}\frac{\partial ^{2}L}{\partial x_{6n+j}\partial x_{3n+i}}%
dx_{6n+j}$

$+X^{6n+i}\frac{\partial ^{2}L}{\partial x_{6n+j}\partial x_{7n+i}}%
dx_{6n+j}-X^{7n+i}\frac{\partial ^{2}L}{\partial x_{6n+j}\partial x_{6n+i}}%
dx_{6n+j}+X^{i}\frac{\partial ^{2}L}{\partial x_{7n+j}\partial x_{n+i}}%
dx_{7n+j}$

$-X^{n+i}\frac{\partial ^{2}L}{\partial x_{7n+j}\partial x_{i}}%
dx_{7n+j}+X^{2n+i}\frac{\partial ^{2}L}{\partial x_{7n+j}\partial x_{4n+i}}%
dx_{7n+j}+X^{3n+i}\frac{\partial ^{2}L}{\partial x_{7n+j}\partial x_{5n+i}}%
dx_{7n+j}$

$-X^{4n+i}\frac{\partial ^{2}L}{\partial x_{7n+j}\partial x_{2n+i}}%
dx_{7n+j}-X^{5n+i}\frac{\partial ^{2}L}{\partial x_{7n+j}\partial x_{3n+i}}%
dx_{7n+j}+X^{6n+i}\frac{\partial ^{2}L}{\partial x_{7n+j}\partial x_{7n+i}}%
dx_{7n+j}$

$-X^{7n+i}\frac{\partial ^{2}L}{\partial x_{7n+j}\partial x_{6n+i}}dx_{7n+j}-%
\frac{\partial L}{\partial x_{j}}dx_{j}-\frac{\partial L}{\partial x_{n+j}}%
dx_{n+j}-\frac{\partial L}{\partial x_{2n+j}}dx_{2n+j}$

$-\frac{\partial L}{\partial x_{3n+j}}dx_{3n+j}-\frac{\partial L}{\partial
x_{4n+j}}dx_{4n+j}-\frac{\partial L}{\partial x_{5n+j}}dx_{5n+j}-\frac{%
\partial L}{\partial x_{6n+j}}dx_{6n+j}-\frac{\partial L}{\partial x_{7n+j}}%
dx_{7n+j}$

With the use of \textbf{Eq.} (\ref{1.1}), the following expressions can be
obtained:

$\ -X^{i}\frac{\partial ^{2}L}{\partial x_{j}\partial x_{n+i}}dx_{j}+X^{i}%
\frac{\partial ^{2}L}{\partial x_{j}\partial x_{i}}dx_{n+j}-X^{i}\frac{%
\partial ^{2}L}{\partial x_{j}\partial x_{4n+i}}dx_{2n+j}-X^{i}\frac{%
\partial ^{2}L}{\partial x_{j}\partial x_{5n+i}}dx_{3n+j}$

$\ +X^{i}\frac{\partial ^{2}L}{\partial x_{j}\partial x_{2n+i}}%
dx_{4n+j}+X^{i}\frac{\partial ^{2}L}{\partial x_{j}\partial x_{3n+i}}%
dx_{5n+j}-X^{i}\frac{\partial ^{2}L}{\partial x_{j}\partial x_{7n+i}}%
dx_{6n+j}$

$+X^{i}\frac{\partial ^{2}L}{\partial x_{j}\partial x_{6n+i}}dx_{7n+j}\ \
-X^{n+i}\frac{\partial ^{2}L}{\partial x_{n+j}\partial x_{n+i}}dx_{j}+X^{n+i}%
\frac{\partial ^{2}L}{\partial x_{n+j}\partial x_{i}}dx_{n+j}$

$-X^{n+i}\frac{\partial ^{2}L}{\partial x_{n+j}\partial x_{4n+i}}dx_{2n+j}\
-X^{n+i}\frac{\partial ^{2}L}{\partial x_{n+j}\partial x_{5n+i}}dx_{3n+j}\
+X^{n+i}\frac{\partial ^{2}L}{\partial x_{n+j}\partial x_{2n+i}}dx_{4n+j}$

$+X^{n+i}\frac{\partial ^{2}L}{\partial x_{n+j}\partial x_{3n+i}}%
dx_{5n+j}-X^{n+i}\frac{\partial ^{2}L}{\partial x_{n+j}\partial x_{7n+i}}%
dx_{6n+j}+X^{n+i}\frac{\partial ^{2}L}{\partial x_{n+j}\partial x_{6n+i}}%
dx_{7n+j}$

$-\ X^{2n+i}\frac{\partial ^{2}L}{\partial x_{2n+j}\partial x_{n+i}}%
dx_{j}+X^{2n+i}\frac{\partial ^{2}L}{\partial x_{2n+j}\partial x_{i}}%
dx_{n+j}-X^{2n+i}\frac{\partial ^{2}L}{\partial x_{2n+j}\partial x_{4n+i}}%
dx_{2n+j}$

$-X^{2n+i}\frac{\partial ^{2}L}{\partial x_{2n+j}\partial x_{5n+i}}%
dx_{3n+j}\ +X^{2n+i}\frac{\partial ^{2}L}{\partial x_{2n+j}\partial x_{2n+i}}%
dx_{4n+j}+X^{2n+i}\frac{\partial ^{2}L}{\partial x_{2n+j}\partial x_{3n+i}}%
dx_{5n+j}$

$-X^{2n+i}\frac{\partial ^{2}L}{\partial x_{2n+j}\partial x_{7n+i}}%
dx_{6n+j}+X^{2n+i}\frac{\partial ^{2}L}{\partial x_{2n+j}\partial x_{6n+i}}%
dx_{7n+j}\ \ -X^{3n+i}\frac{\partial ^{2}L}{\partial x_{3n+j}\partial x_{n+i}%
}dx_{j}$

$+X^{3n+i}\frac{\partial ^{2}L}{\partial x_{3n+j}\partial x_{i}}%
dx_{n+j}-X^{3n+i}\frac{\partial ^{2}L}{\partial x_{3n+j}\partial x_{4n+i}}%
dx_{2n+j}-X^{3n+i}\frac{\partial ^{2}L}{\partial x_{3n+j}\partial x_{5n+i}}%
dx_{3n+j}$

$\ \ +X^{3n+i}\frac{\partial ^{2}L}{\partial x_{3n+j}\partial x_{2n+i}}%
dx_{4n+j}+X^{3n+i}\frac{\partial ^{2}L}{\partial x_{3n+j}\partial x_{3n+i}}%
dx_{5n+j}-X^{3n+i}\frac{\partial ^{2}L}{\partial x_{3n+j}\partial x_{7n+i}}%
dx_{6n+j}$

$+X^{3n+i}\frac{\partial ^{2}L}{\partial x_{3n+j}\partial x_{6n+i}}%
dx_{7n+j}\ \ \ -X^{4n+i}\frac{\partial L}{\partial x_{4n+j}\partial x_{n+i}}%
dx_{j}\ +X^{4n+i}\frac{\partial L}{\partial x_{4n+j}\partial x_{i}}dx_{n+j}$

$-X^{4n+i}\frac{\partial L}{\partial x_{4n+j}\partial x_{4n+i}}dx_{2n+j}\ \
-X^{4n+i}\frac{\partial L}{\partial x_{4n+j}\partial x_{5n+i}}dx_{3n+j}\ \ \
+X^{4n+i}\frac{\partial ^{2}L}{\partial x_{4n+j}\partial x_{2n+i}}dx_{4n+j}$

$+X^{4n+i}\frac{\partial ^{2}L}{\partial x_{4n+j}\partial x_{3n+i}}%
dx_{5n+j}\ -X^{4n+i}\frac{\partial ^{2}L}{\partial x_{4n+j}\partial x_{7n+i}}%
dx_{6n+j}+X^{4n+i}\frac{\partial ^{2}L}{\partial x_{4n+j}\partial x_{6n+i}}%
dx_{7n+j}$

$-X^{5n+i}\frac{\partial ^{2}L}{\partial x_{5n+j}\partial x_{n+i}}dx_{j}\
+X^{5n+i}\frac{\partial ^{2}L}{\partial x_{5n+j}\partial x_{i}}%
dx_{n+j}-X^{5n+i}\frac{\partial ^{2}L}{\partial x_{5n+j}\partial x_{4n+i}}%
dx_{2n+j}$

$-X^{5n+i}\frac{\partial ^{2}L}{\partial x_{5n+j}\partial x_{5n+i}}%
dx_{3n+j}\ \ \ +X^{5n+i}\frac{\partial ^{2}L}{\partial x_{5n+j}\partial
x_{2n+i}}dx_{4n+j}+X^{5n+i}\frac{\partial ^{2}L}{\partial x_{5n+j}\partial
x_{3n+i}}dx_{5n+j}$

$-X^{5n+i}\frac{\partial ^{2}L}{\partial x_{5n+j}\partial x_{7n+i}}%
dx_{6n+j}+X^{5n+i}\frac{\partial ^{2}L}{\partial x_{5n+j}\partial x_{6n+i}}%
dx_{7n+j}-X^{6n+i}\frac{\partial ^{2}L}{\partial x_{6n+j}\partial x_{n+i}}%
dx_{j}$

$+X^{6n+i}\frac{\partial ^{2}L}{\partial x_{6n+j}\partial x_{i}}%
dx_{n+j}-X^{6n+i}\frac{\partial ^{2}L}{\partial x_{6n+j}\partial x_{4n+i}}%
dx_{2n+j}-X^{6n+i}\frac{\partial ^{2}L}{\partial x_{6n+j}\partial x_{5n+i}}%
dx_{3n+j}$

$+X^{6n+i}\frac{\partial ^{2}L}{\partial x_{6n+j}\partial x_{2n+i}}%
dx_{4n+j}\ +X^{6n+i}\frac{\partial ^{2}L}{\partial x_{6n+j}\partial x_{3n+i}}%
dx_{5n+j}-X^{6n+i}\frac{\partial ^{2}L}{\partial x_{6n+j}\partial x_{7n+i}}%
dx_{6n+j}$

$+X^{6n+i}\frac{\partial ^{2}L}{\partial x_{6n+j}\partial x_{6n+i}}%
dx_{7n+j}\ \ \ \ \ -X^{7n+i}\frac{\partial ^{2}L}{\partial x_{7n+j}\partial
x_{n+i}}dx_{j}+X^{7n+i}\frac{\partial ^{2}L}{\partial x_{7n+j}\partial x_{i}}%
dx_{n+j}$

$-X^{7n+i}\frac{\partial ^{2}L}{\partial x_{7n+j}\partial x_{4n+i}}%
dx_{2n+j}-X^{7n+i}\frac{\partial ^{2}L}{\partial x_{7n+j}\partial x_{5n+i}}%
dx_{3n+j}+X^{7n+i}\frac{\partial ^{2}L}{\partial x_{7n+j}\partial x_{2n+i}}%
dx_{4n+j}\ \ $

$+X^{7n+i}\frac{\partial ^{2}L}{\partial x_{7n+j}\partial x_{3n+i}}%
dx_{5n+j}-X^{7n+i}\frac{\partial ^{2}L}{\partial x_{7n+j}\partial x_{7n+i}}%
dx_{6n+j}+X^{7n+i}\frac{\partial ^{2}L}{\partial x_{7n+j}\partial x_{6n+i}}%
dx_{7n+j}$

$+\frac{\partial L}{\partial x_{j}}dx_{j}+\frac{\partial L}{\partial x_{n+j}}%
dx_{n+j}+\frac{\partial L}{\partial x_{2n+j}}dx_{2n+j}+\frac{\partial L}{%
\partial x_{3n+j}}dx_{3n+j}+\frac{\partial L}{\partial x_{4n+j}}dx_{4n+j}$

$+\frac{\partial L}{\partial x_{5n+j}}dx_{5n+j}+\frac{\partial L}{\partial
x_{6n+j}}dx_{6n+j}+\frac{\partial L}{\partial x_{7n+j}}dx_{7n+j}=0$

If a curve denoted by $\alpha :\mathbf{R}\rightarrow \mathbf{R}^{8}$ is
considered to be an integral curve of $\xi ,$ then we calculate the
following equation:

$-X^{i}\frac{\partial ^{2}L}{\partial x_{j}\partial x_{n+i}}dx_{j}\ \
-X^{n+i}\frac{\partial ^{2}L}{\partial x_{n+j}\partial x_{n+i}}dx_{j}-\
X^{2n+i}\frac{\partial ^{2}L}{\partial x_{2n+j}\partial x_{n+i}}dx_{j}\ \ $

$-X^{3n+i}\frac{\partial ^{2}L}{\partial x_{3n+j}\partial x_{n+i}}dx_{j}\
-X^{4n+i}\frac{\partial L}{\partial x_{4n+j}\partial x_{n+i}}dx_{j}\
-X^{5n+i}\frac{\partial ^{2}L}{\partial x_{5n+j}\partial x_{n+i}}dx_{j}$

$-X^{6n+i}\frac{\partial ^{2}L}{\partial x_{6n+j}\partial x_{n+i}}dx_{j}\
-X^{7n+i}\frac{\partial ^{2}L}{\partial x_{7n+j}\partial x_{n+i}}dx_{j}+X^{i}%
\frac{\partial ^{2}L}{\partial x_{j}\partial x_{i}}dx_{n+j}$

$+X^{n+i}\frac{\partial ^{2}L}{\partial x_{n+j}\partial x_{i}}%
dx_{n+j}+X^{2n+i}\frac{\partial ^{2}L}{\partial x_{2n+j}\partial x_{i}}%
dx_{n+j}+X^{3n+i}\frac{\partial ^{2}L}{\partial x_{3n+j}\partial x_{i}}%
dx_{n+j}$

$+X^{4n+i}\frac{\partial L}{\partial x_{4n+j}\partial x_{i}}dx_{n+j}+X^{5n+i}%
\frac{\partial ^{2}L}{\partial x_{5n+j}\partial x_{i}}dx_{n+j}+X^{6n+i}\frac{%
\partial ^{2}L}{\partial x_{6n+j}\partial x_{i}}dx_{n+j}$

$+X^{7n+i}\frac{\partial ^{2}L}{\partial x_{7n+j}\partial x_{i}}%
dx_{n+j}-X^{i}\frac{\partial ^{2}L}{\partial x_{j}\partial x_{4n+i}}%
dx_{2n+j}-X^{n+i}\frac{\partial ^{2}L}{\partial x_{n+j}\partial x_{4n+i}}%
dx_{2n+j}$

$-X^{2n+i}\frac{\partial ^{2}L}{\partial x_{2n+j}\partial x_{4n+i}}%
dx_{2n+j}-X^{3n+i}\frac{\partial ^{2}L}{\partial x_{3n+j}\partial x_{4n+i}}%
dx_{2n+j}-X^{4n+i}\frac{\partial L}{\partial x_{4n+j}\partial x_{4n+i}}%
dx_{2n+j}$

$-X^{5n+i}\frac{\partial ^{2}L}{\partial x_{5n+j}\partial x_{4n+i}}%
dx_{2n+j}-X^{6n+i}\frac{\partial ^{2}L}{\partial x_{6n+j}\partial x_{4n+i}}%
dx_{2n+j}-X^{7n+i}\frac{\partial ^{2}L}{\partial x_{7n+j}\partial x_{4n+i}}%
dx_{2n+j}$

$-X^{i}\frac{\partial ^{2}L}{\partial x_{j}\partial x_{5n+i}}%
dx_{3n+j}-X^{n+i}\frac{\partial ^{2}L}{\partial x_{n+j}\partial x_{5n+i}}%
dx_{3n+j}-X^{2n+i}\frac{\partial ^{2}L}{\partial x_{2n+j}\partial x_{5n+i}}%
dx_{3n+j}$

$-X^{3n+i}\frac{\partial ^{2}L}{\partial x_{3n+j}\partial x_{5n+i}}%
dx_{3n+j}-X^{4n+i}\frac{\partial L}{\partial x_{4n+j}\partial x_{5n+i}}%
dx_{3n+j}-X^{5n+i}\frac{\partial ^{2}L}{\partial x_{5n+j}\partial x_{5n+i}}%
dx_{3n+j}$

$-X^{6n+i}\frac{\partial ^{2}L}{\partial x_{6n+j}\partial x_{5n+i}}%
dx_{3n+j}-X^{7n+i}\frac{\partial ^{2}L}{\partial x_{7n+j}\partial x_{5n+i}}%
dx_{3n+j}+X^{i}\frac{\partial ^{2}L}{\partial x_{j}\partial x_{2n+i}}%
dx_{4n+j}$

$+X^{n+i}\frac{\partial ^{2}L}{\partial x_{n+j}\partial x_{2n+i}}%
dx_{4n+j}+X^{2n+i}\frac{\partial ^{2}L}{\partial x_{2n+j}\partial x_{2n+i}}%
dx_{4n+j}\ +X^{3n+i}\frac{\partial ^{2}L}{\partial x_{3n+j}\partial x_{2n+i}}%
dx_{4n+j}$

$+X^{4n+i}\frac{\partial ^{2}L}{\partial x_{4n+j}\partial x_{2n+i}}%
dx_{4n+j}+X^{5n+i}\frac{\partial ^{2}L}{\partial x_{5n+j}\partial x_{2n+i}}%
dx_{4n+j}+X^{6n+i}\frac{\partial ^{2}L}{\partial x_{6n+j}\partial x_{2n+i}}%
dx_{4n+j}$

$+X^{7n+i}\frac{\partial ^{2}L}{\partial x_{7n+j}\partial x_{2n+i}}%
dx_{4n+j}+X^{i}\frac{\partial ^{2}L}{\partial x_{j}\partial x_{3n+i}}%
dx_{5n+j}+X^{n+i}\frac{\partial ^{2}L}{\partial x_{n+j}\partial x_{3n+i}}%
dx_{5n+j}$

$+X^{2n+i}\frac{\partial ^{2}L}{\partial x_{2n+j}\partial x_{3n+i}}%
dx_{5n+j}+X^{3n+i}\frac{\partial ^{2}L}{\partial x_{3n+j}\partial x_{3n+i}}%
dx_{5n+j}+X^{4n+i}\frac{\partial ^{2}L}{\partial x_{4n+j}\partial x_{3n+i}}%
dx_{5n+j}$

$+X^{5n+i}\frac{\partial ^{2}L}{\partial x_{5n+j}\partial x_{3n+i}}%
dx_{5n+j}+X^{6n+i}\frac{\partial ^{2}L}{\partial x_{6n+j}\partial x_{3n+i}}%
dx_{5n+j}+X^{7n+i}\frac{\partial ^{2}L}{\partial x_{7n+j}\partial x_{3n+i}}%
dx_{5n+j}$

$-X^{i}\frac{\partial ^{2}L}{\partial x_{j}\partial x_{7n+i}}%
dx_{6n+j}-X^{n+i}\frac{\partial ^{2}L}{\partial x_{n+j}\partial x_{7n+i}}%
dx_{6n+j}-X^{2n+i}\frac{\partial ^{2}L}{\partial x_{2n+j}\partial x_{7n+i}}%
dx_{6n+j}$

$-X^{3n+i}\frac{\partial ^{2}L}{\partial x_{3n+j}\partial x_{7n+i}}%
dx_{6n+j}-X^{4n+i}\frac{\partial ^{2}L}{\partial x_{4n+j}\partial x_{7n+i}}%
dx_{6n+j}-X^{5n+i}\frac{\partial ^{2}L}{\partial x_{5n+j}\partial x_{7n+i}}%
dx_{6n+j}$

$-X^{6n+i}\frac{\partial ^{2}L}{\partial x_{6n+j}\partial x_{7n+i}}%
dx_{6n+j}-X^{7n+i}\frac{\partial ^{2}L}{\partial x_{7n+j}\partial x_{7n+i}}%
dx_{6n+j}+X^{i}\frac{\partial ^{2}L}{\partial x_{j}\partial x_{6n+i}}%
dx_{7n+j}$

$+X^{n+i}\frac{\partial ^{2}L}{\partial x_{n+j}\partial x_{6n+i}}%
dx_{7n+j}+X^{2n+i}\frac{\partial ^{2}L}{\partial x_{2n+j}\partial x_{6n+i}}%
dx_{7n+j}+X^{3n+i}\frac{\partial ^{2}L}{\partial x_{3n+j}\partial x_{6n+i}}%
dx_{7n+j}$

$+X^{4n+i}\frac{\partial ^{2}L}{\partial x_{4n+j}\partial x_{6n+i}}%
dx_{7n+j}+X^{5n+i}\frac{\partial ^{2}L}{\partial x_{5n+j}\partial x_{6n+i}}%
dx_{7n+j}+X^{6n+i}\frac{\partial ^{2}L}{\partial x_{6n+j}\partial x_{6n+i}}%
dx_{7n+j}$

$+X^{7n+i}\frac{\partial ^{2}L}{\partial x_{7n+j}\partial x_{6n+i}}dx_{7n+j}+%
\frac{\partial L}{\partial x_{j}}dx_{j}+\frac{\partial L}{\partial x_{n+j}}%
dx_{n+j}+\frac{\partial L}{\partial x_{2n+j}}dx_{2n+j}+\frac{\partial L}{%
\partial x_{3n+j}}dx_{3n+j}$

$+\frac{\partial L}{\partial x_{4n+j}}dx_{4n+j}+\frac{\partial L}{\partial
x_{5n+j}}dx_{5n+j}+\frac{\partial L}{\partial x_{6n+j}}dx_{6n+j}+\frac{%
\partial L}{\partial x_{7n+j}}dx_{7n+j}=0$

\textbf{\ }\ alternatively

$-[X^{i}\frac{\partial ^{2}L}{\partial x_{j}\partial x_{n+i}}\ +X^{n+i}\frac{%
\partial ^{2}L}{\partial x_{n+j}\partial x_{n+i}}+\ X^{2n+i}\frac{\partial
^{2}L}{\partial x_{2n+j}\partial x_{n+i}}+X^{3n+i}\frac{\partial ^{2}L}{%
\partial x_{3n+j}\partial x_{n+i}}\ +X^{4n+i}\frac{\partial L}{\partial
x_{4n+j}\partial x_{n+i}}\ $

$+X^{5n+i}\frac{\partial ^{2}L}{\partial x_{5n+j}\partial x_{n+i}}\ +X^{6n+i}%
\frac{\partial ^{2}L}{\partial x_{6n+j}\partial x_{n+i}}\ +X^{7n+i}\frac{%
\partial ^{2}L}{\partial x_{7n+j}\partial x_{n+i}}]dx_{j}$ $+\frac{\partial L%
}{\partial x_{j}}dx_{j}$

$+[X^{i}\frac{\partial ^{2}L}{\partial x_{j}\partial x_{i}}+X^{n+i}\frac{%
\partial ^{2}L}{\partial x_{n+j}\partial x_{i}}+X^{2n+i}\frac{\partial ^{2}L%
}{\partial x_{2n+j}\partial x_{i}}+X^{3n+i}\frac{\partial ^{2}L}{\partial
x_{3n+j}\partial x_{i}}+X^{4n+i}\frac{\partial L}{\partial x_{4n+j}\partial
x_{i}}$

$+X^{5n+i}\frac{\partial ^{2}L}{\partial x_{5n+j}\partial x_{i}}+X^{6n+i}%
\frac{\partial ^{2}L}{\partial x_{6n+j}\partial x_{i}}+X^{7n+i}\frac{%
\partial ^{2}L}{\partial x_{7n+j}\partial x_{i}}]dx_{n+j}+\frac{\partial L}{%
\partial x_{n+j}}dx_{n+j}$

$-[X^{i}\frac{\partial ^{2}L}{\partial x_{j}\partial x_{4n+i}}+X^{n+i}\frac{%
\partial ^{2}L}{\partial x_{n+j}\partial x_{4n+i}}+X^{2n+i}\frac{\partial
^{2}L}{\partial x_{2n+j}\partial x_{4n+i}}+X^{3n+i}\frac{\partial ^{2}L}{%
\partial x_{3n+j}\partial x_{4n+i}}+X^{4n+i}\frac{\partial L}{\partial
x_{4n+j}\partial x_{4n+i}}$

$+X^{5n+i}\frac{\partial ^{2}L}{\partial x_{5n+j}\partial x_{4n+i}}+X^{6n+i}%
\frac{\partial ^{2}L}{\partial x_{6n+j}\partial x_{4n+i}}+X^{7n+i}\frac{%
\partial ^{2}L}{\partial x_{7n+j}\partial x_{4n+i}}]dx_{2n+j}+\frac{\partial
L}{\partial x_{2n+j}}dx_{2n+j}$

$-[X^{i}\frac{\partial ^{2}L}{\partial x_{j}\partial x_{5n+i}}+X^{n+i}\frac{%
\partial ^{2}L}{\partial x_{n+j}\partial x_{5n+i}}+X^{2n+i}\frac{\partial
^{2}L}{\partial x_{2n+j}\partial x_{5n+i}}+X^{3n+i}\frac{\partial ^{2}L}{%
\partial x_{3n+j}\partial x_{5n+i}}+X^{4n+i}\frac{\partial L}{\partial
x_{4n+j}\partial x_{5n+i}}$

$+X^{5n+i}\frac{\partial ^{2}L}{\partial x_{5n+j}\partial x_{5n+i}}+X^{6n+i}%
\frac{\partial ^{2}L}{\partial x_{6n+j}\partial x_{5n+i}}+X^{7n+i}\frac{%
\partial ^{2}L}{\partial x_{7n+j}\partial x_{5n+i}}]dx_{3n+j}+\frac{\partial
L}{\partial x_{3n+j}}dx_{3n+j}$

$+[X^{i}\frac{\partial ^{2}L}{\partial x_{j}\partial x_{2n+i}}+X^{n+i}\frac{%
\partial ^{2}L}{\partial x_{n+j}\partial x_{2n+i}}+X^{2n+i}\frac{\partial
^{2}L}{\partial x_{2n+j}\partial x_{2n+i}}+X^{3n+i}\frac{\partial ^{2}L}{%
\partial x_{3n+j}\partial x_{2n+i}}+X^{4n+i}\frac{\partial ^{2}L}{\partial
x_{4n+j}\partial x_{2n+i}}$

$+X^{5n+i}\frac{\partial ^{2}L}{\partial x_{5n+j}\partial x_{2n+i}}+X^{6n+i}%
\frac{\partial ^{2}L}{\partial x_{6n+j}\partial x_{2n+i}}+X^{7n+i}\frac{%
\partial ^{2}L}{\partial x_{7n+j}\partial x_{2n+i}}]dx_{4n+j}+\frac{\partial
L}{\partial x_{4n+j}}dx_{4n+j}$

$+[X^{i}\frac{\partial ^{2}L}{\partial x_{j}\partial x_{3n+i}}+X^{n+i}\frac{%
\partial ^{2}L}{\partial x_{n+j}\partial x_{3n+i}}+X^{2n+i}\frac{\partial
^{2}L}{\partial x_{2n+j}\partial x_{3n+i}}+X^{3n+i}\frac{\partial ^{2}L}{%
\partial x_{3n+j}\partial x_{3n+i}}+X^{4n+i}\frac{\partial ^{2}L}{\partial
x_{4n+j}\partial x_{3n+i}}$

$+X^{5n+i}\frac{\partial ^{2}L}{\partial x_{5n+j}\partial x_{3n+i}}+X^{6n+i}%
\frac{\partial ^{2}L}{\partial x_{6n+j}\partial x_{3n+i}}+X^{7n+i}\frac{%
\partial ^{2}L}{\partial x_{7n+j}\partial x_{3n+i}}]dx_{5n+j}+\frac{\partial
L}{\partial x_{5n+j}}dx_{5n+j}$

$-[X^{i}\frac{\partial ^{2}L}{\partial x_{j}\partial x_{7n+i}}+X^{n+i}\frac{%
\partial ^{2}L}{\partial x_{n+j}\partial x_{7n+i}}+X^{2n+i}\frac{\partial
^{2}L}{\partial x_{2n+j}\partial x_{7n+i}}+X^{3n+i}\frac{\partial ^{2}L}{%
\partial x_{3n+j}\partial x_{7n+i}}+X^{4n+i}\frac{\partial ^{2}L}{\partial
x_{4n+j}\partial x_{7n+i}}$

$+X^{5n+i}\frac{\partial ^{2}L}{\partial x_{5n+j}\partial x_{7n+i}}+X^{6n+i}%
\frac{\partial ^{2}L}{\partial x_{6n+j}\partial x_{7n+i}}+X^{7n+i}\frac{%
\partial ^{2}L}{\partial x_{7n+j}\partial x_{7n+i}}]dx_{6n+j}+\frac{\partial
L}{\partial x_{6n+j}}dx_{6n+j}$

$+[X^{i}\frac{\partial ^{2}L}{\partial x_{j}\partial x_{6n+i}}+X^{n+i}\frac{%
\partial ^{2}L}{\partial x_{n+j}\partial x_{6n+i}}+X^{2n+i}\frac{\partial
^{2}L}{\partial x_{2n+j}\partial x_{6n+i}}+X^{3n+i}\frac{\partial ^{2}L}{%
\partial x_{3n+j}\partial x_{6n+i}}+X^{4n+i}\frac{\partial ^{2}L}{\partial
x_{4n+j}\partial x_{6n+i}}$

$+X^{5n+i}\frac{\partial ^{2}L}{\partial x_{5n+j}\partial x_{6n+i}}+X^{6n+i}%
\frac{\partial ^{2}L}{\partial x_{6n+j}\partial x_{6n+i}}+X^{7n+i}\frac{%
\partial ^{2}L}{\partial x_{7n+j}\partial x_{6n+i}}]dx_{7n+j}+\frac{\partial
L}{\partial x_{7n+j}}dx_{7n+j}=0$

Then we obtain the equations

\begin{equation}
\begin{array}{c}
\frac{\partial }{\partial t}\left( \frac{\partial L}{\partial x_{i}}\right) +%
\frac{\partial L}{\partial x_{n+i}}=0,\frac{\partial }{\partial t}\left(
\frac{\partial L}{\partial x_{n+i}}\right) -\frac{\partial L}{\partial x_{i}}%
=0,\frac{\partial }{\partial t}\left( \frac{\partial L}{\partial x_{2n+i}}%
\right) +\frac{\partial L}{\partial x_{4n+i}}=0, \\
\frac{\partial }{\partial t}\left( \frac{\partial L}{\partial x_{3n+i}}%
\right) +\frac{\partial L}{\partial x_{5n+i}}=0,\frac{\partial }{\partial t}%
\left( \frac{\partial L}{\partial x_{4n+i}}\right) -\frac{\partial L}{%
\partial x_{2n+i}}=0,\frac{\partial }{\partial t}\left( \frac{\partial L}{%
\partial x_{5n+i}}\right) -\frac{\partial L}{\partial x_{3n+i}}=0, \\
\frac{\partial }{\partial t}\left( \frac{\partial L}{\partial x_{6n+i}}%
\right) +\frac{\partial L}{\partial x_{7n+i}}=0,\frac{\partial }{\partial t}%
\left( \frac{\partial L}{\partial x_{7n+i}}\right) -\frac{\partial L}{%
\partial x_{6n+i}}=0,%
\end{array}
\label{3.6}
\end{equation}

such that the equations obtained in \textbf{Eq. }(\ref{3.6}) are said to be
\textit{Euler-Lagrange equations} structured on the standard Cliffordian K%
\"{a}hler manifold $(\mathbf{R}^{8},V)$ by means of $\Phi _{L}^{J_{1}}$ and
in the case, the triple $(\mathbf{R}^{8},\Phi _{L}^{J_{1}},\xi )$ is called
a \textit{mechanical system }on the standard Cliffordian K\"{a}hler manifold
$(\mathbf{R}^{8},V)$\textit{.}

Secondly, we find Euler-Lagrange equations for quantum and classical
mechanics by means of $\Phi _{L}^{G}$ on the standard Cliffordian K\"{a}hler
manifold $(M,V).$

Consider $J_{2}$ be another local basis component on the Cliffordian K\"{a}%
hler manifold $(\mathbf{R}^{8},V).$ Let $\xi $ take as in \textbf{Eq.} (\ref%
{3.1}). In the case, the vector field given by

\begin{equation}
\begin{array}{c}
V_{J_{2}}=J_{2}(\xi )=X^{i}\frac{\partial }{\partial x_{2n+i}}-X^{n+i}\frac{%
\partial }{\partial x_{4n+i}}-X^{2n+i}\frac{\partial }{\partial x_{i}}%
+X^{3n+i}\frac{\partial }{\partial x_{6n+i}} \\
+X^{4n+i}\frac{\partial }{\partial x_{n+i}}-X^{5n+i}\frac{\partial }{%
\partial x_{7n+i}}-X^{6n+i}\frac{\partial }{\partial x_{3n+i}}+X^{7n+i}\frac{%
\partial }{\partial x_{5n+i}},%
\end{array}
\label{3.7}
\end{equation}

is \textit{Liouville vector field} on the standard Cliffordian K\"{a}hler
manifold $(\mathbf{R}^{8},V)$. The function given by $%
E_{L}^{J_{2}}=V_{J_{2}}(L)-L$ is\textit{\ energy function}. Then the
operator $i_{J_{2}}$ induced by $J_{2}$ and denoted by%
\begin{equation}
i_{J_{2}}\omega (X_{1},X_{2},...,X_{r})=\sum_{i=1}^{r}\omega
(X_{1},...,J_{2}X_{i},...,X_{r})  \label{3.8}
\end{equation}%
is \textit{vertical derivation, }where $\omega \in \wedge ^{r}{}\mathbf{R}%
^{8},$ $X_{i}\in \chi (\mathbf{R}^{8}).$ The \textit{vertical differentiation%
} $d_{J_{2}}$ are defined by%
\begin{equation}
d_{J_{2}}=[i_{J_{2}},d]=i_{J_{2}}d-di_{J_{2}}.  \label{3.9}
\end{equation}%
Since taking into considering $J_{2},$ the closed standard Clifford K\"{a}%
hler form is the closed 2-form given by $\Phi _{L}^{J_{2}}=-dd_{J_{2}}L$
such that%
\begin{eqnarray}
d_{_{J_{2}}} &=&\frac{\partial }{\partial x_{2n+i}}dx_{i}-\frac{\partial }{%
\partial x_{4n+i}}dx_{n+i}-\frac{\partial }{\partial x_{i}}dx_{2n+i}+\frac{%
\partial }{\partial x_{6n+i}}dx_{3n+i}  \label{3.10} \\
&&+\frac{\partial }{\partial x_{n+i}}dx_{4n+i}-\frac{\partial }{\partial
x_{7n+i}}dx_{5n+i}-\frac{\partial }{\partial x_{3n+i}}dx_{6n+i}+\frac{%
\partial }{\partial x_{5n+i}}dx_{7n+i}  \notag
\end{eqnarray}

and defined by operator%
\begin{equation}
d_{J_{2}}:\mathcal{F}(\mathbf{R}^{8})\rightarrow \wedge ^{1}{}\mathbf{R}^{8}.
\label{3.11}
\end{equation}

The closed standard Clifford K\"{a}hler form $\Phi _{L}^{J_{2}}$ on $\mathbf{%
R}^{8}$ is the symplectic structure. So it holds

\begin{eqnarray}
E_{L}^{J_{2}} &=&V_{J_{2}}(L)-L=X^{i}\frac{\partial L}{\partial x_{2n+i}}%
-X^{n+i}\frac{\partial L}{\partial x_{4n+i}}-X^{2n+i}\frac{\partial L}{%
\partial x_{i}}+X^{3n+i}\frac{\partial L}{\partial x_{6n+i}}  \label{3.12} \\
&&+X^{4n\ +i}\frac{\partial L}{\partial x_{n+i}}-X^{5n+i}\frac{\partial L}{%
\partial x_{7n+i}}-X^{6n+i}\frac{\partial L}{\partial x_{3n+i}}+X^{7n+i}%
\frac{\partial L}{\partial x_{5n+i}}-L  \notag
\end{eqnarray}

By means of \textbf{Eq.} (\ref{1.1}), using (\ref{3.1}), (\ref{3.10}) and (%
\ref{3.12}), also taking into consideration the above first part we
calculate the equations
\begin{equation}
\begin{array}{c}
\frac{\partial }{\partial t}\left( \frac{\partial L}{\partial x_{i}}\right) +%
\frac{\partial L}{\partial x_{2n+i}}=0,\frac{\partial }{\partial t}\left(
\frac{\partial L}{\partial x_{n+i}}\right) -\frac{\partial L}{\partial
x_{4n+i}}=0,\frac{\partial }{\partial t}\left( \frac{\partial L}{\partial
x_{2n+i}}\right) -\frac{\partial L}{\partial x_{i}}=0, \\
\frac{\partial }{\partial t}\left( \frac{\partial L}{\partial x_{3n+i}}%
\right) +\frac{\partial L}{\partial x_{6n+i}}=0,\frac{\partial }{\partial t}%
\left( \frac{\partial L}{\partial x_{4n+i}}\right) +\frac{\partial L}{%
\partial x_{n+i}}=0,\frac{\partial }{\partial t}\left( \frac{\partial L}{%
\partial x_{5n+i}}\right) -\frac{\partial L}{\partial x_{7n+i}}=0, \\
\frac{\partial }{\partial t}\left( \frac{\partial L}{\partial x_{6n+i}}%
\right) -\frac{\partial L}{\partial x_{3n+i}}=0,\frac{\partial }{\partial t}%
\left( \frac{\partial L}{\partial x_{7n+i}}\right) +\frac{\partial L}{%
\partial x_{5n+i}}=0,%
\end{array}
\label{3.13}
\end{equation}%
Hence the equations obtained in \textbf{Eq. }(\ref{3.13}) are called \textit{%
Euler-Lagrange equations} structured by means of $\Phi _{L}^{J_{2}}$ on the
standard Cliffordian K\"{a}hler manifold $(\mathbf{R}^{8},V)$ and so, the
triple $(\mathbf{R}^{8},\Phi _{L}^{J_{2}},\xi )$ is said to be a \textit{%
mechanical system }on the standard Cliffordian K\"{a}hler manifold $(\mathbf{%
R}^{8},V)$\textit{.}

Thirdly, we introduce Euler-Lagrange equations for quantum and classical
mechanics by means of $\Phi _{L}^{J_{3}}$ on the standard Cliffordian K\"{a}%
hler manifold $(\mathbf{R}^{8},V).$

Let $J_{3}$ be a local basis on the standard Cliffordian K\"{a}hler manifold
$(\mathbf{R}^{8},V).$ Let semispray $\xi $ give as in \textbf{Eq.}(\ref{3.1}%
). Therefore, \textit{Liouville vector field} on the standard Cliffordian K%
\"{a}hler manifold $(\mathbf{R}^{8},V)$ is the vector field given by

\begin{equation}
\begin{array}{c}
V_{J_{3}}=J_{3}(\xi )=X^{i}\frac{\partial }{\partial x_{3n+i}}-X^{n+i}\frac{%
\partial }{\partial x_{5n+i}}-X^{2n+i}\frac{\partial }{\partial x_{6n+i}}%
-X^{3n+i}\frac{\partial }{\partial x_{i}} \\
+X^{4n+i}\frac{\partial }{\partial x_{7n+i}}+X^{5n+i}\frac{\partial }{%
\partial x_{n+i}}+X^{6n+i}\frac{\partial }{\partial x_{2n+i}}-X^{7n+i}\frac{%
\partial }{\partial x_{4n+i}}.%
\end{array}
\label{3.14}
\end{equation}%
The function given by $E_{L}^{J_{3}}=V_{J_{3}}(L)-L$ is\textit{\ energy
function} and calculated by
\begin{equation}
\begin{array}{c}
E_{L}^{J_{3}}=X^{i}\frac{\partial L}{\partial x_{3n+i}}-X^{n+i}\frac{%
\partial L}{\partial x_{5n+i}}-X^{2n+i}\frac{\partial L}{\partial x_{6n+i}}%
-X^{3n+i}\frac{\partial L}{\partial x_{i}} \\
+X^{4n+i}\frac{\partial L}{\partial x_{7n+i}}+X^{5n+i}\frac{\partial L}{%
\partial x_{n+i}}+X^{6n+i}\frac{\partial L}{\partial x_{2n+i}}-X^{7n+i}\frac{%
\partial L}{\partial x_{4n+i}}-L.%
\end{array}
\label{3.15}
\end{equation}%
The function $i_{J_{3}}$ induced by $J_{3}$ and shown by%
\begin{equation}
i_{J_{3}}\omega (X_{1},X_{2},...,X_{r})=\sum_{i=1}^{r}\omega
(X_{1},...,J_{3}X_{i},...,X_{r}),  \label{3.16}
\end{equation}%
is said to be \textit{vertical derivation, }where $\omega \in \wedge ^{r}{}%
\mathbf{R}^{8},$ $X_{i}\in \chi (\mathbf{R}^{8}).$ The \textit{vertical
differentiation} $d_{J_{3}}$ is denoted by%
\begin{equation}
d_{J_{3}}=[i_{J_{3}},d]=i_{J_{3}}d-di_{J_{3}},  \label{3.17}
\end{equation}%
Considering $J_{3}$ , the closed K\"{a}hler form is the closed 2-form given
by $\Phi _{L}^{J_{3}}=-dd_{_{J_{3}}}L$ such that

\begin{eqnarray*}
d_{_{J_{3}}} &=&\frac{\partial }{\partial x_{3n+i}}dx_{i}-\frac{\partial }{%
\partial x_{5n+i}}dx_{n+i}-\frac{\partial }{\partial x_{6n+i}}dx_{2n+i}-%
\frac{\partial }{\partial x_{i}}dx_{3n+i} \\
&&+\frac{\partial }{\partial x_{7n+i}}dx_{4n+i}+\frac{\partial }{\partial
x_{n+i}}dx_{5n+i}+\frac{\partial }{\partial x_{2n+i}}dx_{6n+i}-\frac{%
\partial }{\partial x_{4n+i}}dx_{7n+i}
\end{eqnarray*}%
and%
\begin{equation}
d_{_{J_{3}}}:\mathcal{F}(\mathbf{R}^{8})\rightarrow \wedge ^{1}{}\mathbf{R}%
^{8}  \label{3.18}
\end{equation}

Using \textbf{Eq.} (\ref{1.1}), similar to the above first \ and second
cases , we find\ the following expression the equations \
\begin{equation}
\begin{array}{c}
\frac{\partial }{\partial t}\left( \frac{\partial L}{\partial x_{i}}\right) +%
\frac{\partial L}{\partial x_{3n+i}}=0,\frac{\partial }{\partial t}\left(
\frac{\partial L}{\partial x_{n+i}}\right) -\frac{\partial L}{\partial
x_{5n+i}}=0,\frac{\partial }{\partial t}\left( \frac{\partial L}{\partial
x_{2n+i}}\right) -\frac{\partial L}{\partial x_{6n+i}}=0, \\
\frac{\partial }{\partial t}\left( \frac{\partial L}{\partial x_{3n+i}}%
\right) -\frac{\partial L}{\partial x_{i}}=0,\frac{\partial }{\partial t}%
\left( \frac{\partial L}{\partial x_{4n+i}}\right) +\frac{\partial L}{%
\partial x_{7n+i}}=0,\frac{\partial }{\partial t}\left( \frac{\partial L}{%
\partial x_{5n+i}}\right) +\frac{\partial L}{\partial x_{n+i}}=0, \\
\frac{\partial }{\partial t}\left( \frac{\partial L}{\partial x_{6n+i}}%
\right) +\frac{\partial L}{\partial x_{2n+i}}=0,\frac{\partial }{\partial t}%
\left( \frac{\partial L}{\partial x_{7n+i}}\right) -\frac{\partial L}{%
\partial x_{4n+i}}=0,%
\end{array}
\label{3.19}
\end{equation}%
Thus the equations given in \textbf{Eq. }(\ref{3.19}) infer \textit{%
Euler-Lagrange equations} structured by means of $\Phi _{L}^{J_{3}}$ on the
standard Cliffordian K\"{a}hler manifold $(\mathbf{R}^{8},V)$ and therefore
the triple $(\mathbf{R}^{8},\Phi _{L}^{J_{3}},\xi )$ is named a \textit{%
mechanical system }on the standard Cliffordian K\"{a}hler manifold $(\mathbf{%
R}^{8},V)$\textit{.}

\section{Conclusion}

From above, Lagrangian mechanics has intrinsically been described taking
into account a canonical local basis $\{J_{1},J_{2},J_{3}\}$ of $V$ on the
standard Cliffordian K\"{a}hler manifold $(\mathbf{R}^{8},V).$

The paths of semispray $\xi $ on the standard Cliffordian K\"{a}hler
manifold are the solutions Euler--Lagrange equations raised in (\ref{3.6}), (%
\ref{3.13}) and (\ref{3.19}), and also obtained by a canonical local basis $%
\{J_{1},J_{2},J_{3}\}$ of vector bundle $V$ on the standard Cliffordian K%
\"{a}hler manifold $(\mathbf{R}^{8},V)$. \ One can be proved that these
equations are very important to explain the rotational spatial mechanics
problems.

\end{document}